\documentclass[a4paper,12pt]{article}

\usepackage{amsmath}
\usepackage{amssymb}
\usepackage[dvips]{geometry}
\usepackage{fullpage}
\usepackage{helvet}
\usepackage[T1]{fontenc}
\usepackage{graphicx}
\usepackage[amssymb]{SIunits}
\usepackage{url}
\usepackage{citesort}

\newcommand{\J}{\mathrm{j}}  
\newcommand{\D}{\mathrm{d}}  
\newcommand{\E}{\mathrm{e}}  

\newcommand{\mt}{M_{\text{t}}}

\newcommand{\rs}{r^{*}}

\begin{document}

\bibliographystyle{unsrt}

\title{Inversion of spinning sound fields}

\author{Michael Carley,\\
  Department of Mechanical Engineering,\\
  University of Bath,\\
  Bath BA2 7AY,\\
  England;\\
  Electronic mail:~m.j.carley@bath.ac.uk}

\date{\today}


\maketitle


\begin{abstract}

A method is presented for the reconstruction of rotating monopole
source distributions using acoustic pressures measured on a sideline
parallel to the source axis. The method requires no \textit{a priori}
assumptions about the source other than that its strength at the
frequency of interest vary sinusoidally in azimuth on the source disc
so that the radiated acoustic field is composed of a single
circumferential mode. When multiple azimuthal modes are present, the
acoustic field can be decomposed into azimuthal modes and the method
applied to each mode in sequence.

The method proceeds in two stages, first finding an intermediate line
source derived from the source distribution and then inverting this
line source to find the radial variation of source strength. A
far-field form of the radiation integrals is derived, showing that the
far field pressure is a band-limited Fourier transform of the line
source, establishing a limit on the quality of source reconstruction
which can be achieved using far-field measurements. The method is
applied to simulated data representing wind-tunnel testing of a ducted
rotor system (tip Mach number~0.74) and to control of noise from an
automotive cooling fan (tip Mach number~0.14), studies which have
appeared in the literature of source identification.
  
\end{abstract}

\section{Introduction}
\label{sec:intro}

This paper describes a method for determining rotating source
distributions from acoustic measurements. This is a problem which has
been examined by a number of researchers, with
many~\cite{holste-neise97,farassat-nark-thomas01,lewy05,lewy08,%
  castres-joseph07a,castres-joseph07b} considering the problem of
estimating the amplitudes of the acoustic modes at the termination of
a circular duct, as in the case of aircraft engines. The motivation
for these studies has usually been to determine the source terms in
their own right, in order to find the source mechanisms responsible
for the noise or to improve noise control measures, but a second
application has been in developing models which can be used to predict
the acoustic field. This prediction model can be used in active noise
control~\cite{gerard-berry-masson05a,gerard-berry-masson05b} or in
using near-field measurements taken in a wind-tunnel to make far-field
predictions of noise radiated by aircraft in
flight~\cite{holste-neise97,peake-boyd93}. This gives rise to two
different, though related, problems: the first is the determination,
to within some tolerance, of the acoustic source; the second is the
determination of the acoustic source to within a tolerance sufficient
to give accurate predictions of the acoustic field at points other
than the measurement positions.

This paper considers a model problem for the recovery of a rotating
source distribution from a set of measurements along a sideline, a
line parallel to the source axis. The question of how to position
microphones, and how many to use, features in the analysis of many
researchers. Typical microphone configurations have included~3
microphones at~120 angular positions~\cite{holste-neise97}, 91
microphones on a fixed polar array~\cite{castres-joseph07b}, 18
microphones rotating over 20~positions~\cite{farassat-nark-thomas01}
and~21 microphones located on a fixed arc~\cite{lewy08}, depending on
the experimental facilities used and the fidelity of results
required. Recent work on engine noise has also included the use of
sensor arrays mounted inside or on the engine. Examples are the use
of~100 pressure sensors on the surface of the intake\cite{sijtsma07}
and simulations of an array of~150 microphones mounted on the wall of
an engine duct\cite{lowis-joseph06}. In these cases, the methods used
are described as `beamforming' and come from the class of techniques
used for source location rather than for source characterization.

In this paper, we present an inversion technique which uses data from
a linear arrangement of microphones to recover the details of a
distribution of monopoles on a disc. This corresponds to the problem
of thickness noise of a propeller or other rotor~\cite{goldstein74},
sound from a baffled circular piston~\cite{pierce89} or to sound
radiated by the termination of a circular duct, when the Rayleigh
approximation is valid~\cite{tyler-sofrin62,posey-dunn-farassat98}.
The only assumption made is that the source, and the acoustic field,
have a known sinusoidal variation in azimuth---no assumption is
required about the form of the radial variation of the source nor is a
far-field approximation needed. The resulting method is applied to
simulated data using parameters characteristic of problems to which
identification methods have been applied in the past.

The source recovery technique which is developed here is based on the
measurement techniques used in wind tunnel measurement of aerodynamic
sources\cite{holste-neise97,farassat-nark-thomas01,lewy05,lewy08,%
  castres-joseph07a,castres-joseph07b,gerard-berry-masson05a,%
  gerard-berry-masson05b,peake-boyd93} but could also be viewed in the
more general framework of sound source reconstruction in other areas
of acoustics\cite{magalhaes-tenenbaum04} and, in particular, in
relation to cylindrical nearfield acoustical holography
(NAH)\cite{williams99,williams-dardy-washburn87} where measurements
are taken on a cylindrical surface surrounding a source region and
then forward projected to find the acoustic field elsewhere in space,
or back-projected to find the acoustic quantities which characterize
the source. The method of this paper shares some similarities with NAH
but differs in incorporating known information about the source
geometry and azimuthal dependence.

\section{Inversion of spinning sound fields}
\label{sec:inversion}

\begin{figure}
  \centering
  \includegraphics{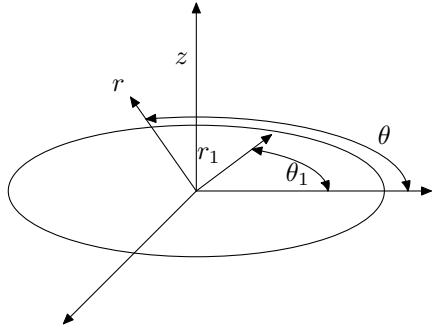}
  \caption{Coordinate system for radiation prediction}
  \label{fig:disc} 
\end{figure}

The acoustic fields to be considered in this paper can all be viewed
as being generated by sources distributed over a
disc. Figure~\ref{fig:disc} shows the arrangement of the problem. We
use cylindrical coordinates $(r,\theta,z)$ with an acoustic monopole
source distributed over the disc, $0\leq r \leq 1$, $z=0$,
non-dimensionalizing all lengths on disc radius. At a single frequency
$\omega$, the acoustic field $p$ is given by~\cite{goldstein74}:
\begin{align}
  \label{equ:full}
  p(r,\theta,z,\omega) &= \int_{0}^{1}\int_{0}^{2\pi}
  s(r_{1},\theta_{1}) \frac{\E^{\J k R}}{4\pi
    R}\,\D\theta_{1}\,r_{1}\,\D r_{1},
\end{align}
with wavenumber $k=\omega/c$, $c$ the speed of sound and
source-observer distance:
\begin{align*}
  R^{2} &= r^{2} + r_{1}^{2} - 2r r_{1}\cos(\theta-\theta_{1}) + z^{2}.
\end{align*}
The source term $s(r_{1},\theta_{1})$ can be decomposed into a series
of azimuthal modes with sinusoidal variation:
\begin{align*}
  s(r_{1},\theta_{1}) &= \sum_{n=-\infty}^{\infty}s_{n}(r_{1})\E^{\J n \theta_{1}},
\end{align*}
which, upon insertion into Equation~\ref{equ:full} with the
transformation $\theta-\theta_{1}\to\theta_{1}$, yields:
\begin{align}
  \label{equ:modes}
  p(r,\theta,z,\omega) &= 
  \sum_{n=-\infty}^{\infty} \E^{\J n \theta}
  \int_{0}^{1}\int_{0}^{2\pi} s_{n}(r_{1})
  \frac{\E^{\J (k R-n\theta_{1})}}{4\pi R}\,\D\theta_{1}\,r_{1}\,\D r_{1},
\end{align}
with $R$ being redefined:
\begin{align*}
  R^{2} &= r^{2} + r_{1}^{2} - 2r r_{1}\cos\theta_{1} + z^{2}.
\end{align*}
The acoustic field at a frequency $\omega$ is thus a sum of azimuthal
modes, each of which is directly generated by a corresponding
azimuthal mode on the source disc. The aim of the inversion procedure
is to recover the source function(s) $s_{n}(r_{1})$ given as input
some acoustic pressures measured in the field. The nature of these
measurements will depend on the type of source being studied.

There are two main categories of problem which will be considered:
rotating sources such as propellers and fans and ducted sources where
the duct termination can be considered a disc-shaped source. For a
source rotating at angular frequency $\Omega$, the radiated field
contains only harmonics of frequency $n\Omega$. Furthermore, if the
source strength is steady in the rotating reference frame, there is
only one azimuthal mode, of order $n$, present at each of these
frequencies. This means that the acoustic field of
Equation~\ref{equ:modes} reduces to:
\begin{align*}
  p(r,\theta,z,n\Omega) &= 
  \E^{\J n \theta}
  \int_{0}^{1}\int_{0}^{2\pi} s_{n}(r_{1})
  \frac{\E^{\J (k R-n\theta_{1})}}{4\pi R}\,\D\theta_{1}\,r_{1}\,\D r_{1}.
\end{align*}

The properties of the acoustic field are largely controlled by the
rotor speed and, in particular, the tip Mach number which, for a
source of unit radius, is $\mt=\Omega/c$. When the source rotates
supersonically, $\mt>1$, the acoustic field is dominated by the source
around the sonic radius $\rs=1/\mt$~\cite{crighton-parry91b}. When
$\mt<1$, the blade tip is the dominant region and, in the far-field,
its radiation is exponentially stronger than that from inboard
regions~\cite{parry-crighton89a}. This means that the measured field
is effectively the field radiated by the tip and recovering the
details of the source at smaller radii will be difficult. On the other
hand, if the aim is to accurately compute the acoustic field at a new
set of points, it may well be sufficient to capture only the source at
the tip.

The structure of the rotating field has been studied using model
solutions~\cite{chapman93,carley99,carley00} and the role of the sonic
radius has been clarified. The field is made up of a segmented near
field which undergoes a transition around $\rs$. In `tunneling' across
this transition region, the sound field decays exponentially,
explaining the relatively weak field radiated by subsonically rotating
sources. Supersonic sources have part of the source lying beyond $\rs$
so that they can radiate strongly into the field, without losing
energy in tunneling through the transition. This transition region
means that source recovery will always be a hard problem if only
far-field data are available, a result which will be derived in
\S\ref{sec:far:field} by considering the bandwidth of the spatial data
in the far field.

When the radiating system is a circular duct, the problem can be
modeled by taking the noise source to be the duct termination. In that
case, the source distribution is composed of the duct modes which have
propagated to the end of the duct. The field inside a circular duct is
composed of modes of the form $J_{n}(k_{mn}r)\exp[\J
(n\theta-k_{zmn}z)]$, with $J_{n}(\cdot)$ the Bessel function of the
first kind, $J'_{n}(k_{mn})=0$ and $k_{zmn}$ is an axial
wavenumber~\cite{tyler-sofrin62}. When $k_{zmn}$ has an imaginary
part, the mode decays exponentially in the duct and does not propagate
to the termination. In any case, the source strength at the duct
termination can be taken to be the acoustic velocity generated by the
modes which do propagate and the radiated noise can be accurately
computed over much of the field using a Rayleigh
integral~\cite{tyler-sofrin62,lewy05} or a Kirchhoff integral over a
wider range of polar angles~\cite{lewy05,hocter99}. The source can
again be modeled as a circular disc with an azimuthally varying source
term. A number of methods have been developed for the identification
of the radiating modes~\cite{farassat-nark-thomas01,lewy05,lewy08,%
  castres-joseph07a,castres-joseph07b} and have been found to be
accurate and robust, considering the assumptions made in their
development. 

The one extra difficulty in the duct case compared to the rotor
problem is that the source at the duct termination may be composed of
modes of more than one azimuthal order. In this case, there are
procedures which use measurements at multiple angles to extract the
modal amplitudes in the acoustic field. For example, a method has been
presented which uses~360 measurements distributed over a semicircular
`hoop' to find the amplitudes of the azimuthal modes radiated from a
duct\cite{farassat-nark-thomas01}. The hoop of microphones was then
moved to find the modal amplitudes as a function of axial displacement
$z$.

From the known properties of rotating acoustic fields and established
experimental techniques, it is clear that it is possible to measure
and/or extract the complex amplitude of a single azimuthal mode
radiated by a disc-shaped source. Indeed, if the source is tonal so
that the modal content does not change with time, the measurements
could, in principle, be performed with only two microphones, one fixed
as a phase reference, and another moving along the sideline.

\subsection{Formulation}
\label{sec:formulation}

\begin{figure}
  \centering
  \includegraphics{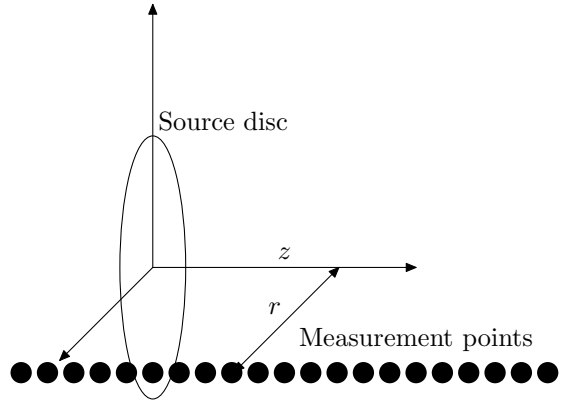}
  \caption{Arrangement of experimental measurements} 
  \label{fig:setup} 
\end{figure}

Figure~\ref{fig:setup} shows the basic experimental arrangement.  The
input to the inversion method is the amplitude of a single azimuthal
mode $p(r,z)$ with $r$ fixed. When the sound is generated by a steady
rotating source, $p(r,z)$ can be found by measuring the field on one
sideline. When modes of different azimuthal order are present, the
field must be measured on multiple sidelines of the same radius $r$,
varying $\theta$, and a decomposition procedure applied to find
$p(r,z)$, as discussed in the previous section.



However the acoustic field may have been measured and processed, the
sound radiated by one source mode of azimuthal order $n$ at frequency
$\omega$ is found by integration over the source
disc~\cite{goldstein74}:
\begin{align}
  \label{equ:basic}
  p(r,z) &= \int_{0}^{1}f(r_{1})\int_{0}^{2\pi}
  \frac{\E^{\J(kR-n\theta_{1})}}{4\pi R}\,\D\theta_{1}r_{1}\,\D r_{1},
\end{align}
where the observer is positioned at $(r,0,z)$. The aim of the
inversion algorithm is to recover the radial source distribution
$f(r_{1})$ from the field pressures $p(r,z)$. 

\begin{figure}
  \centering
  \includegraphics{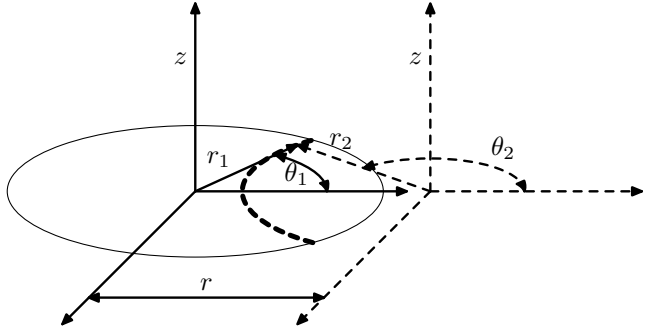}
  \caption{Coordinate systems $(r_{1},\theta_{1},z)$ and
    $(r_{2},\theta_{2},z)$. The $(r_{2},\theta_{2},z)$ system is shown
    dashed and the thick line shows the region of integration over
    $\theta_{2}$ in the transformed system.} \label{fig:coordinates}
\end{figure}

To begin to recover $f(r_{1})$, the first stage is to rewrite
Equation~\ref{equ:basic} in a transformed coordinate system
$(r_{2},\theta_{2},z)$ centred on the measurement sideline,
Figure~\ref{fig:coordinates}. This transformation has been used in
calculations of transient radiation from
pistons~\cite{oberhettinger61a,pierce89} and in studies of propeller
noise fields~\cite{chapman93,carley99,carley00,carley01}. Transforming
Equation~\ref{equ:basic} gives $p(r,z)$ as an integral over a line
source $K(r,r_{2})$:
\begin{align}
  \label{equ:transformed}
  p(r,z) &= \int_{r-1}^{r+1} \frac{\E^{\J kR}}{R}
  K(r,r_{2})r_{2}\,\D
  r_{2},\\
  R &= \left(r_{2}^{2} + z^{2}\right)^{1/2},\nonumber\\
  \label{equ:kfunc}
  K(r,r_{2}) &= \frac{1}{4\pi}
  \int_{\theta_{2}^{(0)}}^{2\pi-\theta_{2}^{(0)}} \E^{-\J
    n\theta_{1}}f(r_{1})\,\D\theta_{2},
\end{align}
for observer positions with $r>1$. The original coordinates 
$(r_{1},\theta_{1})$ are related to $(r_{2},\theta_{2})$ by:
\begin{subequations}
  \label{equ:reverse}
  \begin{align}
    r_{1}^{2} &= r^{2} + r_{2}^{2} + 2rr_{2}\cos\theta_{2},\\
    \theta_{1} &= \tan^{-1}
    \frac{r_{2}\sin\theta_{2}}{r+r_{2}\cos\theta_{2}}.
  \end{align}
\end{subequations}
so that the limits of integration in Equation~\ref{equ:kfunc} are
given by setting $r_{1}=1$:
\begin{align}
  \label{equ:theta}
  \theta_{2}^{(0)} &= \cos^{-1}\frac{1-r^{2}-r_{2}^{2}}{2rr_{2}}.
\end{align}

The function $K(r,r_{2})$ depends only on the observer lateral
displacement and is constant for all points on a sideline parallel to
the source axis. The inversion method proposed is to measure $p(r,z)$
at fixed $r$, invert Equation~\ref{equ:transformed} to recover
$K(r,r_{2})$ and then use Equation~\ref{equ:kfunc} to recover
$f(r_{1})$.

\subsection{Inversion algorithm}
\label{sec:algorithm}

The first stage of the inversion procedure is to use measured sideline
data to recover the source function $K(r,r_{2})$. Noting the behavior
of $K$ at its endpoints, Equation~\ref{equ:kfunc:ends}, we write:
\begin{align}
  \label{equ:kfunc:root}
  K(r,r_{2}) &= [(r_{2}-(r-1))(r+1-r_{2})]^{1/2}K'(r,r_{2}).
\end{align}
The integral of Equation~\ref{equ:transformed} is discretized to give:
\begin{align}
  \label{equ:system:1}
  \sum_{i=1}^{N}\frac{\E^{\J k R_{ij}}}{R_{ij}}
    (r+t_{i}^{(N)})w_{i}^{(N)} K_{i}' &= p_{j},
\end{align}
where:
\begin{align*}
  R_{ij} &= 
  \left[
    (r+t_{i}^{(N)})^{2} + z_{j}^{2}
  \right]^{1/2},
\end{align*}
$z_{j}$ is the axial displacement of the $j$th measurement point and
$(t_{i}^{(N)},w_{i}^{(N)})$ are the nodes and weights of an $N$-point
Gauss-Chebyshev quadrature rule of the second kind.

Equation~\ref{equ:system:1} can be written as a system of equations
relating the vector of measured pressures $\mathbf{p}$ to the unknown
vector of sources $\mathbf{K'}$:
\begin{align}
  \label{equ:system:2}
  [\mathbf{A}] \mathbf{K'} &= \mathbf{p},\\
  A_{ji} &= \frac{\E^{\J k R_{ij}}}{R_{ij}} (r+t_{i}^{(N)})w_{i}^{(N)}.
\end{align}
In practice, the system will be over-determined, with the number of
measured pressures $M$ being greater than $N$, the number of values of
$K'$ to be determined. At this stage, the system is solved for
$\mathbf{K'}$, using some suitable method for ill-conditioned
problems, with $K$ being recovered from Equation~\ref{equ:kfunc:root}.

The second stage in determining the source distribution is to invert
Equation~\ref{equ:kfunc} to recover $f(r_{1})$. We proceed by
approximating $f(r_{1})$ as a sum of Legendre polynomials
$P_{q}(r_{1})$:
\begin{align}
  \label{equ:func:poly}
  f(r_{1}) &= \sum_{q=0}^{Q} F_{q}P_{q}(r_{1}),
\end{align}
so that
\begin{align}
  \label{equ:kfunc:poly}
  K(r,r_{2}) &= \frac{1}{4\pi}
  \sum_{q=0}^{Q} F_{q}
  \int_{\theta_{2}^{(0)}}^{2\pi-\theta_{2}^{(0)}} \E^{-\J
    n\theta_{1}}P_{q}(r_{1})\,\D\theta_{2},  
\end{align}
giving rise to the system of equations:
\begin{align}
  \label{equ:system:kfunc}
  [\mathbf{B}]\mathbf{F} &= \mathbf{K},\\
  B_{iq} &= \frac{1}{4\pi}
  \int_{\theta_{2}^{(0)}}^{2\pi-\theta_{2}^{(0)}} 
  \E^{-\J n\theta_{1}}P_{q}(r_{1})\,\D\theta_{2},\\
  r_{2} &= r+t_{i}^{(N)}. \nonumber
\end{align}
The integration is performed using a standard Gauss-Legendre
quadrature. As before, this system can be solved using a method
suitable for ill-conditioned problems and $f(r_{1})$ reconstructed
from the coefficient vector $\mathbf{F}$.

\subsection{Far-field limitations}
\label{sec:far:field}

The integral of Equation~\ref{equ:transformed} is identical to the
exact integral of Equation~\ref{equ:basic}. If we make the standard
far-field approximations, we can establish some limit on the accuracy
of reconstruction possible using far field results. Expanding $R$ to
first order in $r_{2}$:
\begin{align*}
  R &\approx R_{0} + \frac{r}{R_{0}}(r_{2}-r),\quad R_{0} =
  \left[r^{2}+z^{2}\right]^{1/2},
\end{align*}
so that:
\begin{align}
  \label{equ:far:field}
  p & \approx \frac{\E^{\J k(R_{0}-r^{2}/R_{0})}}{R_{0}}
    \int_{r-1}^{r+1} \E^{\J k r r_{2}/R_{0}} K(r,r_{2})r_{2}\,\D r_{2},
\end{align}
which can be rewritten:
\begin{align}
  \label{equ:fourier}
  p &\approx \frac{\E^{\J k(R_{0}-r^{2}/R_{0})}}{R_{0}}
  \int_{-\infty}^{\infty} \E^{\J\alpha r_{2}} K(r,r_{2})r_{2}
  H(r_{2}-(r-1))H(r+1-r_{2})\,\D r_{2},\\
  \alpha &= kr/R_{0},\nonumber
\end{align}
where $H(\cdot)$ is the Heaviside step function~\cite{lighthill58}.

In the far field, Equation~\ref{equ:fourier} shows that the measured
pressure on a sideline is proportional to a band-limited Fourier
transform of the source term $K(r,r_{2})$. A reconstruction algorithm
based on far field measurements can only recover components of
$K(r,r_{2})$ with spatial frequency $0\leq\alpha\leq k$, with
components outside this frequency band being lost in tunneling across
the transition region between the near and far fields. 

This result provides a link between NAH and the method of this
paper. In NAH, a Fourier transform on the sideline data,
i.e. Equation~\ref{equ:fourier}, is used to recover the coefficients
of a field expansion in cylindrical
wave-functions\cite{williams-dardy-washburn87,williams99}.
This leads to difficulties with finite aperture effects due to the
periodicity enforced by the finite Fourier transform. In this
algorithm, no use is made of the Fourier transform in the
reconstruction procedure so that the shortcomings of the finite
Fourier transform do not cause the spurious sources which appear in
NAH. On the other hand, the discretization introduced by the finite
number of samples on the sideline can lead to aliasing as in NAH and
any other reconstruction procedure based on spatial sampling of the
acoustic field. The implication of Equation~\ref{equ:fourier} is that
in order to use the information which is present on the sideline, the
sampling rate must be such as to capture behavior up to wavenumber
$k$. If the minimum sampling rate is taken to be twice per wavelength
then the sideline measurements should be taken no more than $\pi/k$
apart.

\section{Results}
\label{sec:results}

Two test cases have been simulated as a first test of the algorithm of
\S\ref{sec:algorithm}. The first uses parameters characteristic of the
CRISP ducted rotor tests\cite{holste-neise97} and the second models an
automotive cooling fan which has been used in tests of noise
control\cite{gerard-berry-masson05a,gerard-berry-masson05b}. In each
case, the sound field $p(z)$ is computed by integration of
Equation~\ref{equ:basic}. To simulate measurement errors and
background noise, a Gaussian random signal of amplitude
$\epsilon\max|p|$ is added to the computed pressures before using them
in the inversion scheme.

For the ducted rotor test case, $\mt=0.74$, $k=7.4$, $n=10$, $M=128$,
$N=64$, $r=1.125$ and $0\leq z\leq 4$. The source $f(r_{1})$ was
synthesized by adding the first four duct modes of circumferential
order $n$ with a random phase so that the source was given by:
\begin{align*}
  f(r_{1}) &= \sum_{m=1}^{4} \E^{\J\phi_{m}}J_{n}(k_{mn}r_{1}),
\end{align*}
where $\phi_{m}$ is a random phase $0\leq\phi_{m}\leq2\pi$. The number
of measurement points $N$ was chosen to be approximately equal to that
in the CRISP tests where data were taken at~120
points\cite{holste-neise97}.

In the cooling fan case, $\mt=0.14$, $k=0.83$, $n=6$, $M=16$, $N=16$,
$r=1.25$ and $0\leq z\leq 8$. This time, the source used was
$f(r_{1})=(1-r_{1})^{1/2}$, as this gives reasonable physical
behaviour near the blade tip\cite{parry-crighton89a}. Again, the
number of sensor positions was chosen to be similar to that used in
the original work: in this case, 17 microphones were used in the
authors' source reconstruction
experiments\cite{gerard-berry-masson05a,gerard-berry-masson05b}. 

The two test cases which have been chosen represent two realistic
problems with quite different characteristics. The CRISP case is
similar to many wind tunnel tests which aim to extract the acoustic
source from in-field measurements: the source is quite high frequency
and the tip Mach number is such that, although energy is lost in the
transition to the far field, the acoustic field is quite strong and
there is sufficient information to allow the source to be determined
reasonably accurately. The low-speed cooling fan, however, presents a
rather more difficult problem. Due to the low rotor speed, the field
decays rapidly inside the sonic radius $\rs=7.14$ and the measured
field has lost much of the content useful for source reconstruction. 

The inversion method has been implemented using Octave\cite{octave}
and the Regularization Tools package of
Hansen\cite{hansen94,hansen07}. Equation~\ref{equ:system:2} is solved
using Hansen's implementation of truncated singular value
decomposition~\cite{hansen08} with the regularization parameter
automatically selected using the L-curve
criterion~\cite{hansen-oleary93}. The same technique is then used to
solve Equation~\ref{equ:system:kfunc} and find $f(r_{1})$.

Two measures are used to assess the accuracy of the method. The first
is to compare the recovered source $g(r_{1})$ with the input
$f(r_{1})$. The second is to use $g(r_{1})$ to compute the acoustic
field $q(r,z)$ at a new set of points and compare this field to
$p(r,z)$ computed using $f(r_{1})$. This assesses the ability of the
algorithm to `project' measured data into the field.

\subsection{Source reconstruction}
\label{sec:reconstruction}

\begin{figure}
  \centering
  \includegraphics{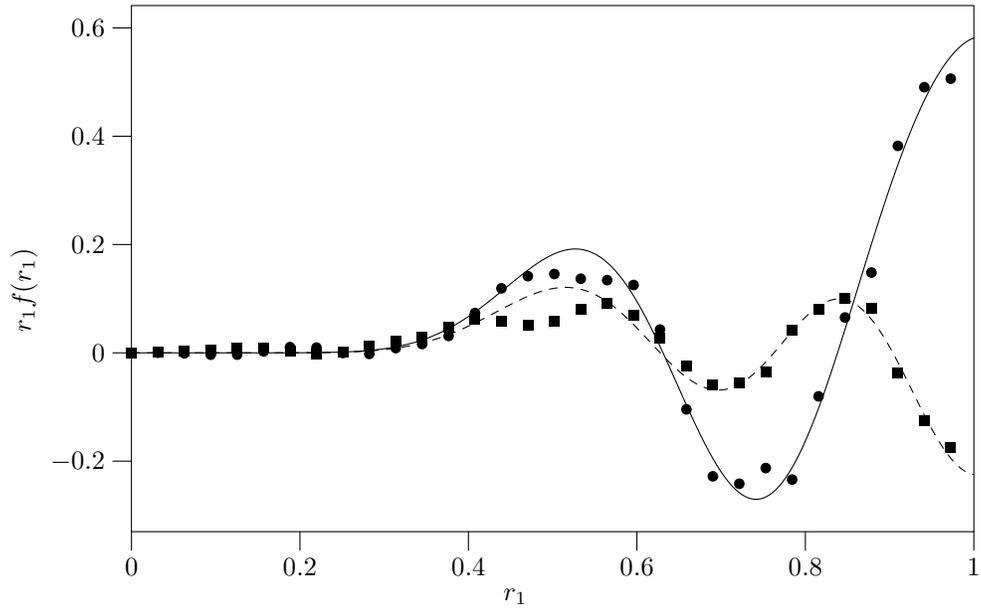}
  \caption{Ducted fan test case, source term, $\epsilon=0$; solid and
    dashed lines, $\Re(f)$ and $\Im(f)$; circles and squares, $\Re(g)$
    and $\Im(g)$.}
\label{fig:holste:1} 
\end{figure}

\begin{figure}
  \centering
  \includegraphics{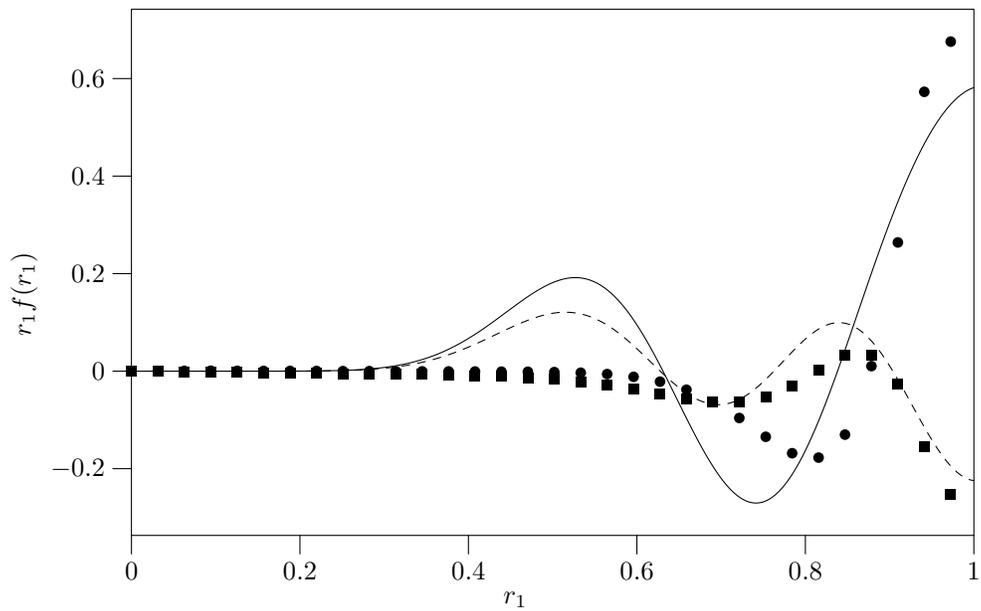}
  \caption{Ducted fan test case, source term,
    $\epsilon=10^{-3}$.} \label{fig:holste:3}
\end{figure}

\begin{figure}
  \centering
  \includegraphics{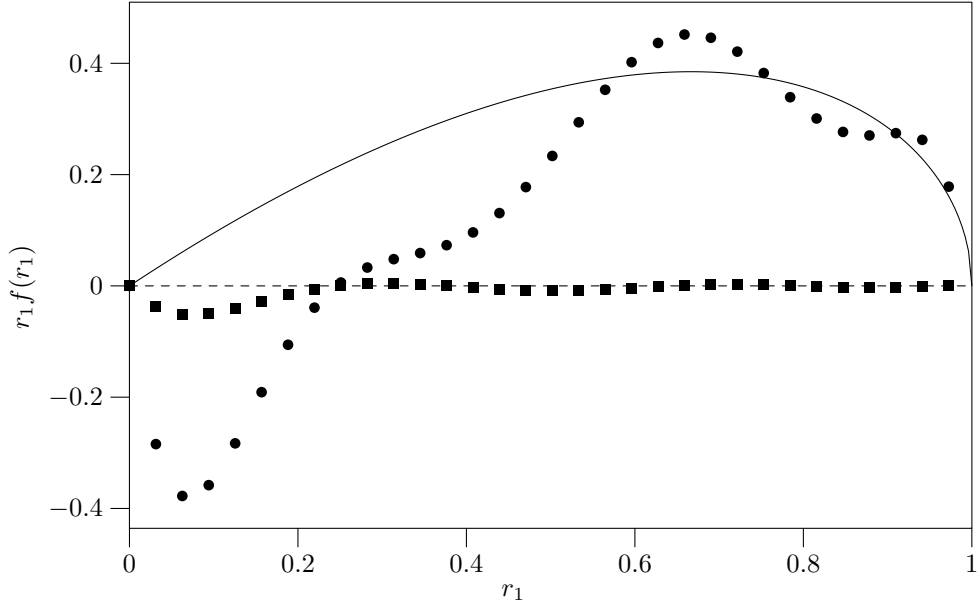}
  \caption{Cooling fan test case, source term,
    $\epsilon=0$.} \label{fig:gerard:1}
\end{figure}

\begin{figure}
  \centering
  \includegraphics{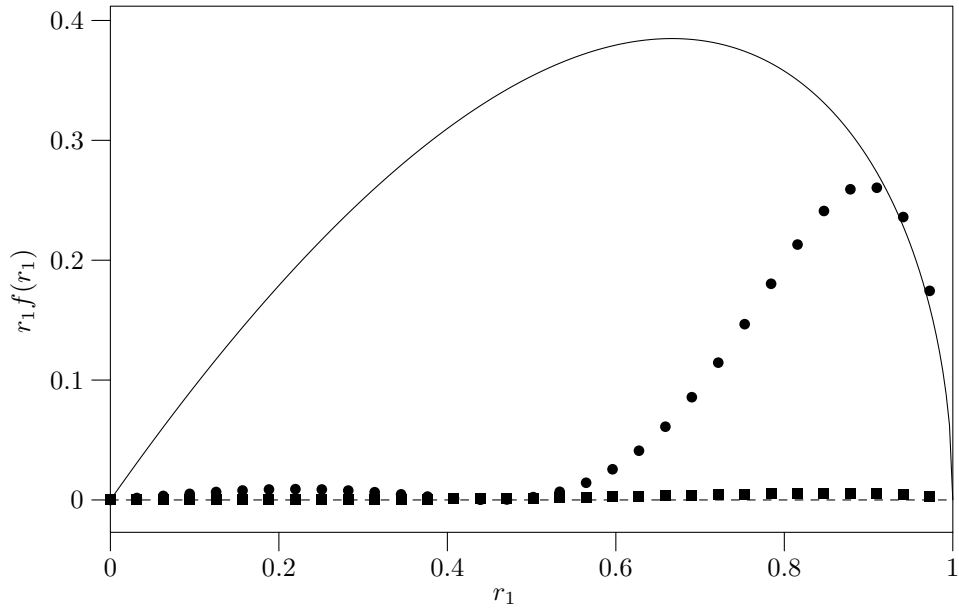}
  \caption{Cooling fan test case, source term,
    $\epsilon=10^{-3}$.} \label{fig:gerard:3}
\end{figure}

The inversion algorithm has been run with zero added noise and with
$\epsilon=10^{-3}$, equivalent to a maximum signal-to-noise ratio of
60\deci\bel. Figures~\ref{fig:holste:1}--\ref{fig:gerard:3} show the
reconstructed source for the two test cases, with the source terms
weighted on radius $r_{1}$, as in the radiation integrals. The source
reconstruction in the ducted fan case Figures~\ref{fig:holste:1}
and~\ref{fig:holste:3} is quite good in both cases. With zero noise,
it accurately reproduces the shape and amplitude of the input
source. With added noise, the reconstruction is not quite as good,
especially for inboard $r_{1}\lessapprox0.8$ but the details of the
source are captured quite well near $r_{1}=1$, the dominant region for
radiation at this wavenumber.

The cooling fan results, Figures~\ref{fig:gerard:1}
and~\ref{fig:gerard:3} are not as good, probably because the number of
sensors is quite small but also because the acoustic field is so much
weaker than in the ducted fan case, due to the low rotor speed. As
discussed in \S\ref{sec:inversion}, sound from source regions inside
the sonic radius decays exponentially as it radiates. Here the whole
source lies inside the sonic radius, meaning that the acoustic field
is composed largely of evanescent waves, making source reconstruction
difficult.

In Figure~\ref{fig:gerard:1}, the reconstructed source oscillates
considerably at small radii, but the tip behaviour is very well
captured. This might be expected: the tip is strongly dominant meaning
that the recovery of the inboard source is very poorly
conditioned. With noise added, the reconstructed source is smoother,
although the amplitude is not found accurately. The tip behaviour,
however, is again accurately computed.

\subsection{Field estimation}
\label{sec:estimation}

\begin{figure}
  \centering
  \includegraphics{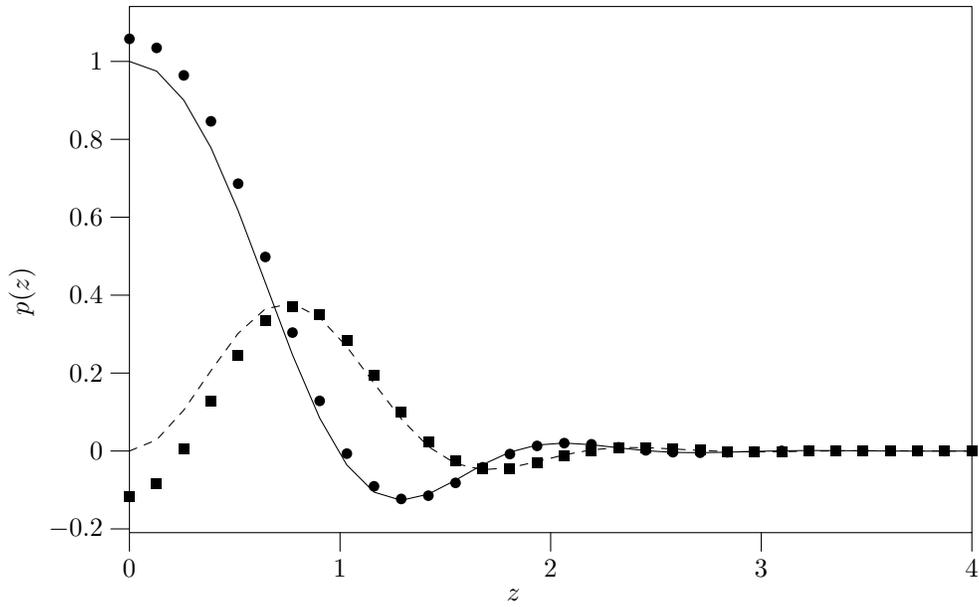}
  \caption{Ducted fan test case, reconstructed near-field noise,
    $\epsilon=10^{-3}$: solid and dashed lines, $\Re(p)$ and $\Im(p)$;
    circles and squares, $\Re(q)$ and
    $\Im(q)$.} \label{fig:holste:n:3}
\end{figure}

\begin{figure}
  \centering
  \includegraphics{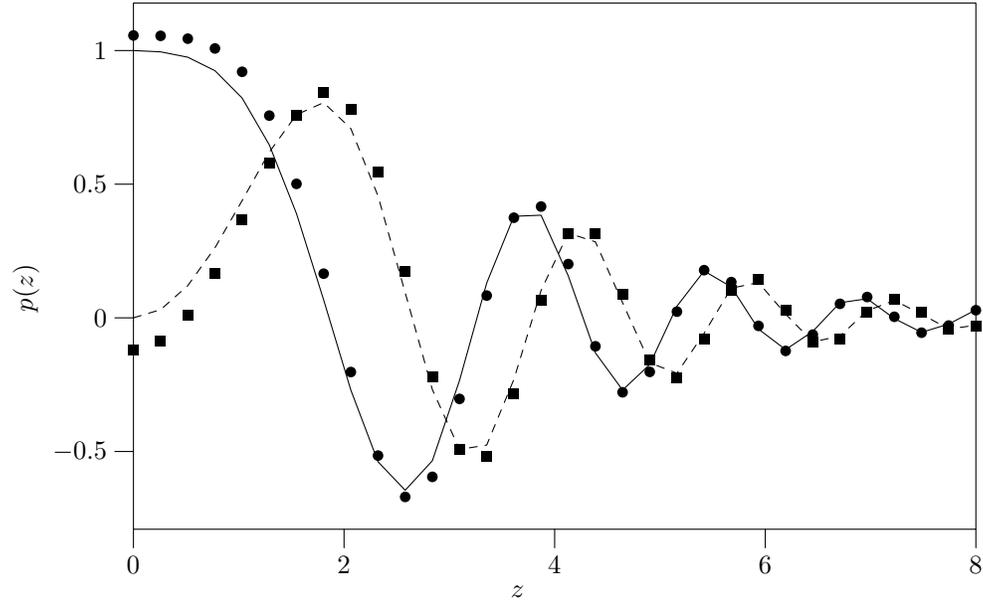}
  \caption{Ducted fan test case, reconstructed far-field noise,
    $\epsilon=10^{-3}$.} \label{fig:holste:f:3}
\end{figure}

\begin{figure}
  \centering
  \includegraphics{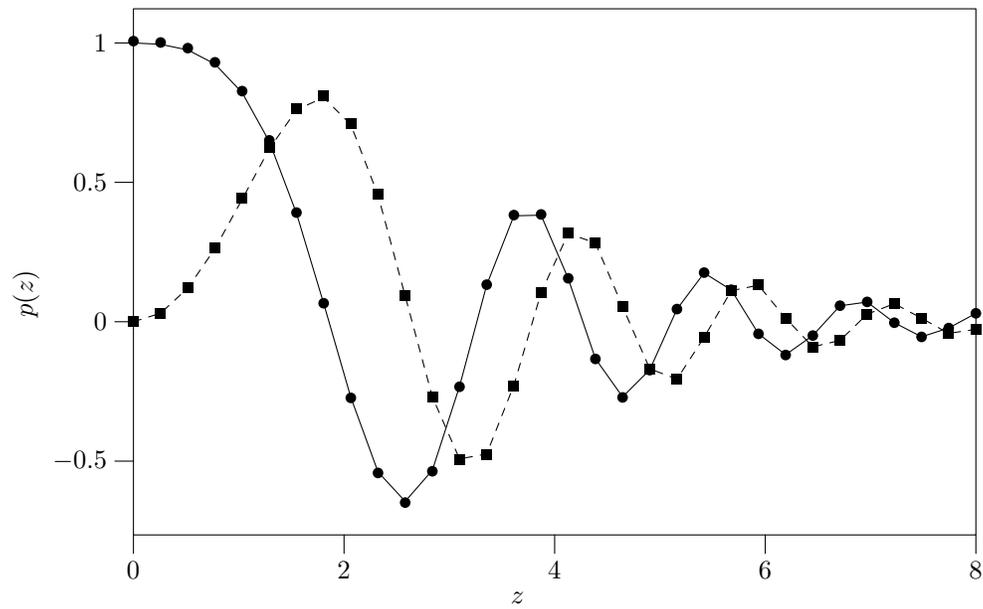}
  \caption{Ducted fan test case, reconstructed far-field noise,
    $\epsilon=0$.} \label{fig:holste:f:1}
\end{figure}

\begin{figure}
  \centering
  \includegraphics{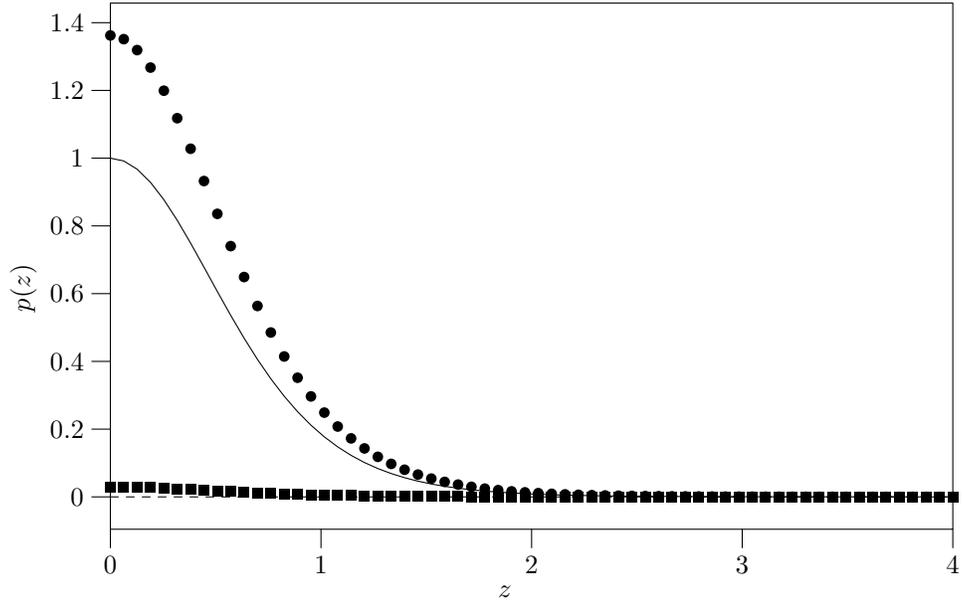}
  \caption{Cooling fan test case, reconstructed near-field noise,
    $\epsilon=10^{-3}$.} \label{fig:gerard:n:3}
\end{figure}

\begin{figure}
  \centering
  \includegraphics{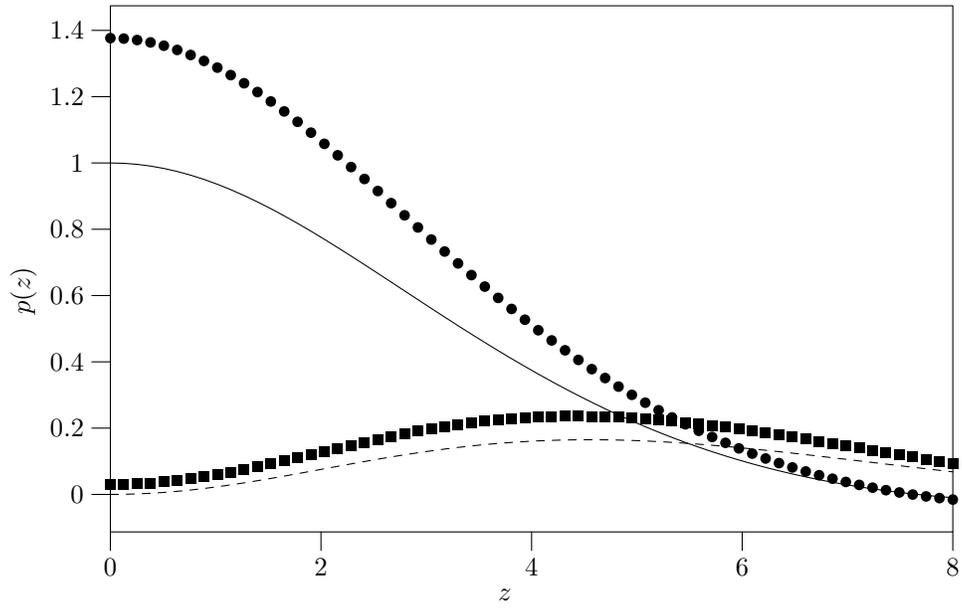}
  \caption{Cooling fan test case, reconstructed far-field noise,
    $\epsilon=10^{-3}$.} \label{fig:gerard:f:3}
\end{figure}

Figures~\ref{fig:holste:n:3}--\ref{fig:gerard:f:3} compare the field
computed using $g(r_{1})$ to the real field $p(r,z)$, near to ($r=2$)
and far from ($r=8$) the source disc, for the $\epsilon=10^{-3}$
case. The results have been scaled on $p(r,0)$ to simplify
comparison. Real and imaginary parts are shown separately as a check
on the ability of the method to calculate the phase of the field,
important in scattering calculations and in control.

The ducted fan results, Figures~\ref{fig:holste:n:3}
and~\ref{fig:holste:f:3}, are very good. The phase has been accurately
computed in the near and far fields and the amplitude error is about
10\% of peak amplitude, or~1\deci\bel. The directivity of the source
is such that the field does not decay rapidly on the sideline, aiding
the reconstruction technique. As a check that the method does converge
to a correct result in the absence of noise, the reconstruction method
has also been applied to data with $\epsilon=0$. The recomputed
far-field pressures are shown in Figure~\ref{fig:holste:f:1} and, as
they should be, are very close to the correct data, indeed practically
indistinguishable from them with the amplitude error at $z=0$ being
0.06\deci\bel. 

In the cooling fan case, Figures~\ref{fig:gerard:n:3}
and~\ref{fig:gerard:f:3}, the field decays rapidly and is
reconstructed quite poorly.The shape and phase are roughly correct but
the amplitude error is about~50\% or~4\deci\bel. The error may be due
to the form of the field or to the small number of sensors
simulated. Note that although the amplitude of the reconstructed
source is much less than that of $f(r_{1})$, the reconstructed field
amplitude is rather larger. This is due to the exponential dominance
of the tip region as an acoustic source on subsonic rotors, mentioned
in \S\ref{sec:inversion}: the difference in tip gradient for
$g(r_{1})$ has made the computed acoustic field stronger than that
found using $f(r_{1})$.

\begin{figure}
  \centering
  \includegraphics{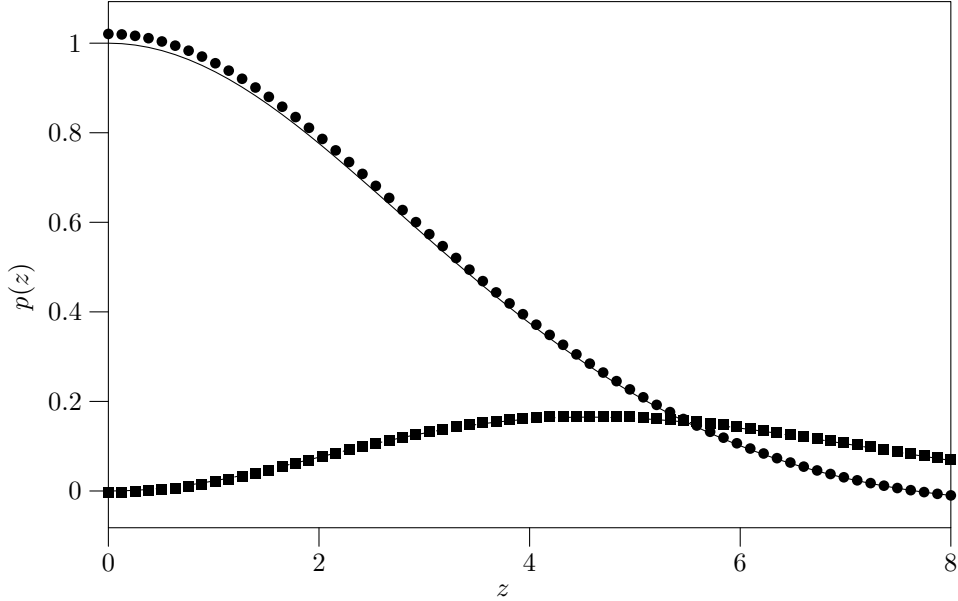}
  \caption{Cooling fan test case, reconstructed far-field noise,
    $\epsilon=0$.} \label{fig:gerard:f:1}
\end{figure}

In any case, given that the phase has been accurately computed, the
result might still be useful in control applications where the phase
of the control signal is important in cancelling the unwanted
noise. Again, we present results with no added noise,
Figure~\ref{fig:gerard:f:1}, and, here, the comparison is not as good
as in the ducted rotor case. The shape of the field has been well
captured but the amplitude is underestimated by about 3\% or
0.26\deci\bel.

\subsection{Algorithm performance}
\label{sec:performance}

\begin{figure}
  \centering
  \includegraphics{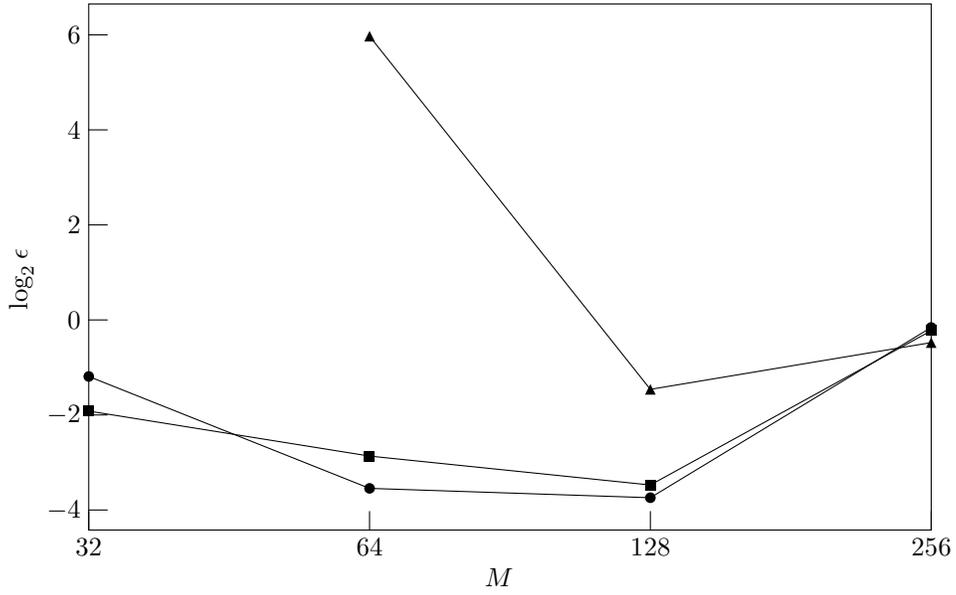}
  \caption{Error $\epsilon$ versus number of sensor positions $M$ and
    $M/N$: squares: $M=N$; circles: $M=2N$; triangles:
    $M=4N$.} \label{fig:nmodes}
\end{figure}

\begin{figure}
  \centering
  \includegraphics{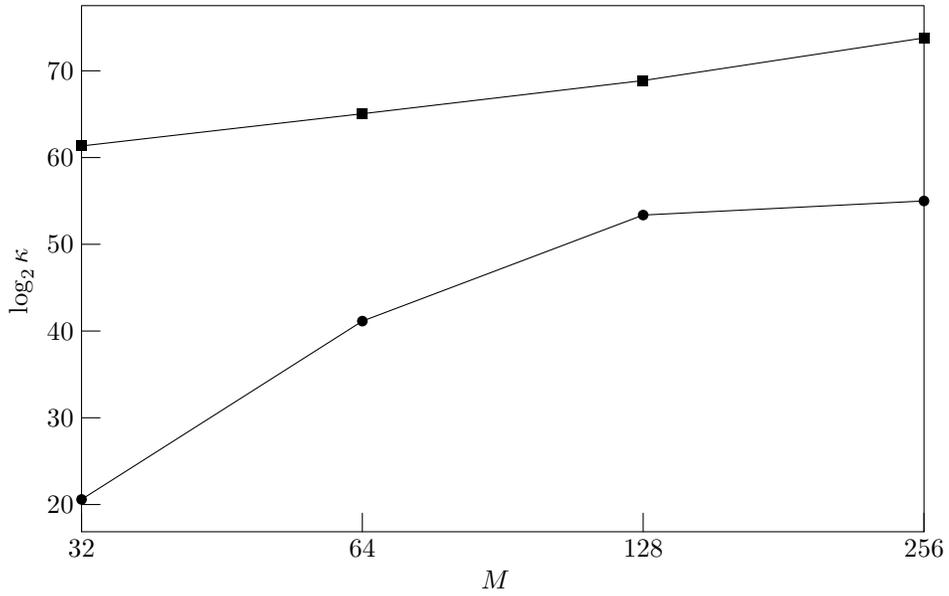}
  \caption{Condition number $\kappa$ of reconstruction matrices versus
    $M$, $N=M$: squares: $[\mathbf{A}]$; circles:
    $[\mathbf{B}]$.} \label{fig:condition}
\end{figure}

To assess the performance of the method when used with varying numbers
of measurements, the source reconstruction method has been applied to
the simulated CRISP data with $M=16, 64, 128, 256$ and
$M/N=1,2,4$. The error measure for the reconstructed source is the
$L_{\infty}$ norm:
\begin{align*}
  L_{\infty} =
  \frac{\max|r_{1}f(r_{1})-r_{1}g(r_{1})|}{\max|r_{1}(f(r_{1})|},
\end{align*}
where the weighting with $r_{1}$ has been retained, corresponding to
an area weighting of the error. The calculation has been performed
with no added noise, to check the factors which contribute to the
error in the reconstructed quantities. 

Figure~\ref{fig:nmodes} shows the variation of error with the number
of sensors, as a function of the ratio $M/N$. The first obvious point
is that, for this set of operating parameters, the error for $M=64$ is
very large when $M/N=4$, while the method failed completely at
$M=32$. This appears to indicate that the source term cannot be
well-approximated by only~16 terms in the expansion of
Equation~\ref{equ:func:poly}, an unsurprising result.

More interesting is that for $N/M=1,2$, the error decreases steadily
as $M$ increases, but then increases between $M=128$ and
$M=256$. Figure~\ref{fig:condition} shows the condition number
$\kappa$ of the matrices $[\mathbf{A}]$ and $[\mathbf{B}]$ used in the
inversion procedure, as a function of $M$, with $N=M$. As might be
expected, the condition number of both increases with $M$, as the
systems become more poorly conditioned. Machine precision on the
computer used for the calculations is approximately $1/2^{52}$. The
condition number $\kappa(\mathbf{A})$ of the matrix used to estimate
$K(r,r_{2})$ is always greater than $2^{60}$ so that the first part of
the inversion scheme is always ill-conditioned. The reason for the
drop in accuracy past $M=128$ seems to be the higher condition number
of the second matrix used $\kappa(\mathbf{B})$. At $M=128$, it rises
above $2^{52}$ and we conjecture that at this point the loss of
precision in calculations is too great for the inversion method and
the results begin to worsen. This is a function of the solver used and
it may be that a different choice of regularization scheme would lead
to better results.


\section{Conclusions}
\label{sec:conclusions}

A source reconstruction method for the inversion of spinning acoustic
fields has been developed and tested on two representative
problems. It has been found that the method can work well, even with
added noise, depending on the type of source to be identified. The
method requires no \textit{a priori} assumptions about the form of the
source other than that it be circular and vary sinusoidally in
azimuth. This makes it a useful intermediate between nearfield
acoustical holography, where no information is assumed about the
source except its approximate location, and other source
identification methods which use assumptions about the location and
spatial variation of the source to model its radiation
characteristics. In the case of the method of this paper, it may be
that when additional information about the source is available, such
as its modal structure, users might be able to incorporate this
information into the technique to improve source reconstruction and/or
to reduce the number of measurements required.

\appendix

\section{End-point behaviour of $K(r,r_{2})$}
\label{sec:endpoint}

To establish the behavior of $K(r,r_{2})$ near the endpoints of the
integrand in Equation~\ref{equ:transformed}, we note that as
$r_{2}\to(r-1)$, $\theta_{2}^{(0)}\to\pi$ and $\theta_{1}\to0$. When
$r_{2}\to(r+1)$, $\theta_{2}^{(0)}\to\pi$ and $\theta_{1}\to\pi$. We
examine the basic integral:
\begin{align}
  \label{equ:kfunc:basic}
  K &= \frac{1}{4\pi} \int_{\theta_{2}^{(0)}}^{2\pi-\theta_{2}^{(0)}}
  \E^{-\J n\theta_{1}}\,\D\theta_{2}.
\end{align}
For $\theta_{2}^{(0)}\to\pi$ and resulting small $\theta_{1}$:
\begin{align}
  \label{equ:kfunc:approx}
  K &\approx \frac{1}{4\pi} 
  \int_{\theta_{2}^{(0)}}^{2\pi-\theta_{2}^{(0)}} 1 \,\D\theta_{2}.
\end{align}
Integrating,
\begin{align*}
  K &\approx \left(2\pi-2\theta_{2}^{(0)}\right)/4\pi
\end{align*}
and using $\cos^{-1} x\to(1-x^{2})^{1/2}$ as $x\to-1$ to give:
\begin{align*}
  \theta_{2}^{(0)} &\approx
  \pi-\frac{2^{1/2}(r+1-r_{2})^{1/2}(r_{2}-(r-1))^{1/2}}
  {\left(2rr_{2}\right)^{1/2}},
\end{align*}
yields
\begin{align}
  \label{equ:kfunc:ends}
  K &\approx \frac{(1+r-r_{2})^{1/2}(r_{2}-(r-1))^{1/2}}{2(rr_{2})^{1/2}},
\end{align}
with square root behaviour as $r_{2}\to (r-1)^{+}$ and $r_{2}\to
(r+1)^{-}$.

\bibliography{abbrev,duct,solid,maths,aerodynamics,scattering,misc,%
  propnoise,identification}




\end{document}